\documentclass[aps,prb,twocolumn,groupedaddress,showpacs,showkeys]{revtex4}

\usepackage{amsfonts}
\usepackage{graphicx}
\def \figwidth{8.5cm}
\begin{document}

\title{Study of coupling loss on bi-columnar BSCCO/Ag tapes by a.c. susceptibility measurements}
\author{D. Zola}\thanks{Corresponding author}
\email[\newline e-mail:]{zoldan@sa.infn.it}
\thanks{\newline FAX: +3908965275 \\}
\affiliation{Department of Physics \lq\lq E. R. Caianiello\rq\rq ,
and INFM Research Unit, University of Salerno, via S. Allende,
I-84081 Baronissi, (Salerno), Italy. }
\author{M. Polichetti}
\affiliation{Department of Physics \lq\lq E. R. Caianiello\rq\rq ,
and INFM Research Unit, University of Salerno, via S. Allende,
I-84081 Baronissi, (Salerno), Italy. }
\author{S. Pace}
\affiliation{Department of Physics \lq\lq E. R. Caianiello\rq\rq ,
and INFM Research Unit, University of Salerno, via S. Allende,
I-84081 Baronissi, (Salerno), Italy. }

\author{F. G\"om\"ory}
 \affiliation{Institute  of  Electrical  Engineering,\
 Slovak  Academy  of  Sciences, \ Dubravska  Cesta  9 \ 842 39  Bratislava,  Slovakia }

\author{F. Str\'{y}\v{c}ek}
\affiliation{Institute  of  Electrical  Engineering,\
 Slovak  Academy  of  Sciences, \ Dubravska  Cesta  9 \ 842 39  Bratislava,  Slovakia }

\author{E. Seiler}
\affiliation{Institute  of  Electrical  Engineering,\
 Slovak  Academy  of  Sciences, \ Dubravska  Cesta  9 \ 842 39  Bratislava,  Slovakia }

\author{I. Hu\v{s}ek}
 \affiliation{Institute  of  Electrical  Engineering,\
 Slovak  Academy  of  Sciences, \ Dubravska  Cesta  9 \ 842 39  Bratislava,  Slovakia }

\author{P. Kov\'{a}\v{c}}
 \affiliation{Institute  of  Electrical  Engineering,\
 Slovak  Academy  of  Sciences, \ Dubravska  Cesta  9 \ 842 39  Bratislava,  Slovakia }

\date{\today}

\begin{abstract}
Coupling losses were studied in composite tapes containing
superconducting material in the form of two separate stacks of
densely packed filaments embedded in a metallic matrix of Ag or Ag
alloy. This kind of sample geometry is quite favorable for
studying the coupling currents and in particular the role of
superconducting bridges between filaments. By using a.c.
susceptibility technique, the electromagnetic losses as function
of a.c. magnetic field amplitude and frequency were measured at
the temperature $T = 77$~K for two tapes with different matrix
composition. The length of samples was varied by subsequent
cutting in order to investigate its influence on the dynamics of
magnetic flux penetration. The geometrical factor $\chi_0$ which
takes into account the demagnetizing effects was established from
a.c. susceptibility data at low amplitudes. Losses vs frequency
dependencies have been found to agree nicely with the theoretical
model developed for round multifilamentary wires. Applying this
model, the effective resistivity of the matrix was determined for
each tape, by using only measured quantities. For the tape with
pure silver matrix its value was found to be larger than what
predicted by the theory for given metal resistivity and
filamentary architecture. On the contrary, in the sample with a
Ag/Mg alloy matrix, an effective resistivity much lower than
expected was determined. We explain these discrepancies by taking
into account the properties of the electrical contact of the
interface between the superconducting filaments and the normal
matrix. In the case of soft matrix of pure Ag, this  is of poor
quality, while the properties of alloy matrix seem to provoke an
extensive creation of intergrowths which can be actually observed
in this kind of samples.
\end{abstract}


\maketitle


\section{\label{Introduction}Introduction}
BSCCO/Ag tapes usually contain many superconducting filaments
embedded in a silver or silver alloy matrix. For achieving thermal
stability and quench protection, the metallic matrix with high
thermal conductivity should have a good interface with the
superconducting material. When the tape is immersed in alternating
(a.c.) magnetic field, an energy loss is observed that is caused
by various mechanisms\cite{SST10(1997)733,Clem}. The irreversible
magnetisation is an intrinsic property of type II superconductors,
and the energy loss related to it is usually called "hysteretic
loss". The a.c. magnetic field induces also an electrical field
which drives currents in the loops formed by pairs of filaments
and closed by crossing the metallic matrix. The ohmic loss
generated in the metallic part of the loop is accordingly called
"coupling loss". Finally, the currents, which flow in metallic
matrix only, give rise to the so called "eddy currents loss". The
latter contribution can be often omitted at the power grid
frequency.  In general, it is desirable to take measures that
reduce the different kinds of losses. To reduce the hysteretic
losses, the dimensions of superconducting filaments have to be
diminished. To cut down the coupling losses, it is essential to
reduce the area of induced flux e.g. by twisting the filaments.
Moreover, the matrix resistivity should rise, for example by
manufacturing an artificial resistive barrier around the
filaments. The effective resistivity of the paths that the
coupling currents follow generally differs from the bare matrix
resistivity, because it depends on the tape geometry and on the
particular arrangement of the superconducting filaments. It is
often used to characterize the coupling currents by the time
constant $\tau$ of magnetic flux penetration. As shown by
Kwasnitza\cite{Cryo17(1977)617}, the coupling loss can be
expressed as a function of $\omega \tau$ ($\omega = 2\pi\nu$) at
constant a.c. magnetic field amplitude ($B_0$). In other words,
the coupling loss is well characterized by this time constant. For
example, in an application working at frequency $\nu$ the tape
should be used that exhibits the time constant such that
$2\pi\nu\tau \ll 1$ to avoid the saturation of the filaments just
by the induced coupling currents. Existing theories provide
formulas for the time constant in a tape with known geometry and
matrix resistivity. Unfortunately, in HTS tapes the experimental
data on the time constant can be very different from the
calculated value because of the irregularities in the interface
between the filaments and the matrix. Therefore it is necessary to
understand how the coupling loss works in real BSCCO/Ag tapes.
Moreover it is necessary to understand what is the
 metallic matrix which has the best effective resistivity.
\\ \indent Thorough investigations have been carried out on
coupling loss in HTS tapes
\cite{PC233(1994)423,PC290(1997)281,PC299(1998)113,Cryo39(1999)829,PC337(2000)187,PC355(2001)325,PC370(2002)177}.
However, it is still not evident whether the coupling mechanism in
BSCCO/Ag can be analyzed in the same manner as for low T$_c$
wires. In this paper, this kind of losses has been studied in
tapes with a geometry resembling just two filaments separated by a
metallic matrix. In reality, each "filament" consists in dense
stack of extremely flat filaments. In this way a simple geometry
has been  achieved, allowing us to study the coupling currents
without other spurious effects. We have measured the time constant
and determined the effective resistivity for two such bi-columnar
tapes that differ significantly in the matrix composition and thus
in its resistivity. The external magnetic field ($B_a$) was always
perpendicular to the broad face of the sample. The geometrical
factor $\chi_0$ (related to demagnetizing factor)
\cite{PRB61(2000)6413,SST13(2000)1327} has been measured. In order
to study the effect of the sample length on the losses, the
measurements were repeated for the same tape, cut several times in
always shorter pieces. In this way we could discuss the various
aspects of coupling losses in this system and  make a comparison
with the most used model\cite{Cryo22(1982)3} in this field. The
importance of the intergrowths between the superconducting
filaments on the coupling losses is also pointed out.

\section{\label{sec2}Theoretical background}
The coupling loss  per unit volume and per cycle($Q_c$) depends
mainly on $\omega\tau$ and on the square of the a.c. field
amplitude ($B_0$). Several authors give expressions of $Q_c$ for
harmonic external fields, perpendicular to the broad face of the
tape
\cite{Cryo17(1977)617,IEEETM13(1977)524,Cryo22(1982)3,PC233(1994)423}.
In our analysis the Campbell  model has been
used\cite{Cryo22(1982)3}. The model assumes that the
superconductor is divided into many filaments and the magnetic
field ($B$) inside the composite conductor, averaged on many
filaments, is uniform. From this last requirement it follows that
the coupling currents  flow in the outer shell  of the wire
\cite{Cryo22(1982)3}. Since these currents  are proportional to
the time derivative ($\dot{B}$) of internal magnetic field $B$, we
can write the relation
\begin{equation}\label{eq2}
  B = B_a - \tau\dot{B}
\end{equation}
where $B_a$ is the applied magnetic field.
 The proportionality constant $\tau$ is
called \emph{time constant} of the tape and the theory states that
it is related to the sample geometry (the cross section as well as
the length) and to the matrix resistivity. $\tau$ is also the time
decay of the coupling currents if the external magnetic field
$B_a$ becomes instantaneously constant. For this reason, in
principle, the time constant could be measured by magnetic
relaxation measurements. Nevertheless these measurements are
influenced by the relaxation of the irreversible magnetisation of
the superconducting filaments and by the overshoot of the magnetic
field produced by the magnet \cite{unp}. By integrating the
Eq.~(\ref{eq2}) it is possible to calculate the coupling loss
density ($Q_c)$ for a sinusoidal external field:
\begin{equation}\label{eq3}
  Q_c =
  \frac{B_0^2}{2\mu_0}\left[2\pi\chi_0\left(\frac{\omega\tau}{1+\omega^2\tau^2}\right)\right
  ]
\end{equation}
where $\chi_0$ is the demagnetization factor in the actual sample
vs. field geometry. The expression in square brackets is
independent of magnetic field amplitude. Moreover we can suppose
that the frequency dependence predicted by eq. (\ref{eq3}) is
still valid also in presence of  few filaments. In this case we
can expect that the eq. (\ref{eq3}) has to be corrected by a
numerical factor:
\begin{equation}\label{eq3a}
  Q_c' = \gamma Q_c
 \end{equation}
On the other hand the a.c. losses are related to the imaginary
part of the first harmonic of the a.c. susceptibility
\cite{Clem,PRB61(2000)6413} and can be expressed as:
\begin{equation}\label{eq1}
  Q = \pi\chi''\chi_0B_0^2/\mu_0
\end{equation}
where $\chi''$ is the imaginary part of the a.c. susceptibility.
By supposing that the losses are dominated by coupling mechanism
it holds:
\begin{equation}\label{eq3b}
   \chi'' = \gamma \frac{\omega\tau}{1+\omega^2\tau^2}
\end{equation}
 By measuring the loss as a function of frequency
($\nu$) at fixed $B_0$, the time constant can be determined by
finding the frequency ($\nu_m$) where the maximum occurs,
corresponding to $\omega \tau = 1 $. On the other hand, in the low
frequency limit $Q_c$ depends linearly on $\tau$ and therefore,
$\tau$ can be evaluated by measuring magnetisation loops for
different sweep rates of a small external field
\cite{PC335(2000)164}. Nevertheless the linear dependence has to
be separated from the logarithmic dependence due to the flux creep
in the filaments \cite{PC372-376(2002)1823}, which has not been
considered in the Campbell model.\\ \indent The
equation~(\ref{eq3}) is  no more valid when the filaments become
saturated by currents, i.e. the current density is equal to the
critical current density in the whole section of the filament.
 In fact for $B_0$ values larger than the full
penetration field of a single superconducting filament ($B_{pf}$),
or as $\omega B_0$ becomes quite large,  the currents saturate the
outer filaments. Then the filaments behave as fully coupled and
the coupling loss turns to be  hysteretic. Since the coupling
currents propagate inside the specimen, the field inside the
sample is no more uniform, which was the key assumption for the
validity of the equation~(\ref{eq3}). On the other hand, for
increasing $B_0$, the magnetic moment of the sample rises.
Nevertheless, it cannot exceed the magnetic moment of an
equivalent superconducting tape with a single core offering the
identical cross section and the same $J_c$ of the multifilamentary
tape. Therefore at high field, the losses of a multifilamentary
tape can be found from the area of the hysteresis loop of the
equivalent monocore tape. This upper limit of loss can be exactly
evaluated by taking into account the sample geometry.
Nevertheless, for $B_0 \gg B_{ps}$ (where $B_{ps}$ is the magnetic
field where the saturation in the magnetisation occurs) one can
neglect the actual shape of $M(B_a)$ loop and the extremities of
$M(B_a)$ loops can be approximate by upright straight lines. Thus,
the saturation loss is written as:
\begin{equation}\label{eq3c}
  Q_c = 4 \mu_0 M_s B_0
\end{equation}
where $M_s$ is the saturation magnetisation. \\ \indent As far as
the frequency dependence is concerned, for $B_0 \gg B_{ps}$, as
the frequency increases, the losses saturate at a constant
value\cite{Cryo22(1982)3} given by eq. (\ref{eq3c}) which is, in a
first approximation, frequency independent.  Also in this case,
going more insight, $Q_c$ should be considered as frequency
dependent due to the frequency dependence of $M_s$ which is not
always negligible in high T$_c$ superconductors
\cite{PC372-376(2002)1823}.

\subsection{\label{sec2:1} Expressions for time constant}
Theoretical expressions for $\tau$  can be found by calculating
the electrical field $E$ inside the tape and  the power loss
density $p = E^2 /\rho_{m}$, where $\rho_m$ is the effective
resistivity of the metallic matrix. At low frequencies this
quantity can be compared to the expression given by the Campbell
model\cite{Cryo22(1982)3}:
\begin{equation}\label{eq3d}
  {p}=\chi_{0}\tau\frac{\dot{B^{2}}}{2\mu_{0}}
\end{equation}
to extract $\tau$. In literature, expressions for $\tau$ are
reported \cite{PC355(2001)325}, for twisted and untwisted flat
cable, for external magnetic field either perpendicular or
parallel to the broad face of the tape. In our case, the samples
are  untwisted BSCCO tapes and the expression for $\tau$ is:
\begin{equation}\label{eq4}
  \tau =\frac{\ell^{2}\mu_0}{\pi^{2}\chi_{0}\rho_{m}}
\end{equation}
showing that the time constant depends on the square of the
length, on the resistivity and on the tape geometry (represented
by $\chi_0$ factor).\\ If $\chi_0$ is 1 (the value for a slab
geometry) the previous expression is identical to the expression
for $\tau$ found solving the flux diffusion equation in a slab.

\subsection{\label{sec2:2}Effective resistivity}
\begin{figure}[htb]
\includegraphics[width=\figwidth, clip]{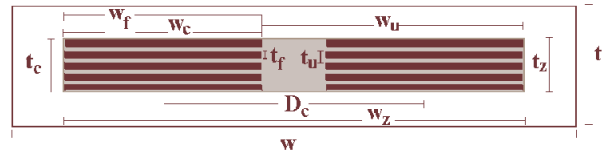}
\caption{Schematic cross section of the tape. $w$ and $t$ are
respectively the width and the thickness of the tape. $w_z$ and
$t_z$ are the dimensions of the filamentary zone. $w_c$ and $t_c$
are respectively the width and the thickness of the two columns.
$t_f$ is the filament thickness, whereas $w_f$ is the filament
width. $w_u$ is the column width plus the distance between the two
columns, whereas $t_u$ is the filament thickness plus the distance
between the neighbor filaments. Finally $D_c$ is the distance
between the columns centers.} \label{disegnotape}
\end{figure}

In the expression (\ref{eq4}), the effective resistivity depends
also on the arrangement of the filaments in the tape. Carr
\cite{Carr83} has shown that, in the continuum anisotropic model,
the effective resistivity is defined in terms of the matrix
resistivity $\rho_m$ and the superconducting volume fraction
$\eta_{eff}$ related to the filamentary zone alone.  In BSCCO tape
$\eta_{eff}$ is larger than the superconducting fraction $\eta$
due to a presence of an external metallic sheath that embeds the
filaments. In the particular case of a round wire, and supposing
that the electrical contact between the metallic matrix and the
superconducting filaments is good, the effective resistivity is
lower than $\rho_m$ and it is given by \cite{Carr83}:
\begin{equation}\label{eq5}
  \rho_{eff}=\rho_{m}\frac{1-\eta_{eff}}{1+\eta_{eff}}
\end{equation}
 If the
electrical contact is bad, giving a high resistive barrier around
each filament,  $\rho_{eff}$ increases in comparison with the
resistivity of the matrix, according to the expression
\cite{Carr83}:
\begin{equation}\label{eq6}
  \rho_{eff}=\rho_{m}\frac{1+\eta_{eff}}{1-\eta_{eff}}
\end{equation}
\indent In our sample, there was no artificially involved
resistive barrier . We also assume that heat treatments commonly
increases the quality of the electrical contact between filaments
and matrix. Modification of the expressions (\ref{eq5}) and
(\ref{eq6}), taking into account geometries more similar to our
samples, have been calculated \cite{SST13(2000)1101}, yelding for
the geometry of our samples the ratio $\rho_{eff}/\rho_m $ as:
\begin{equation}\label{eq7}
\frac{\rho_{eff}}{\rho_m}=\frac{t_uw_u-t_uw_f}{t_uw_u -
t_uw_f+t_fw_f}
\end{equation}
where the meaning of $t_u$, $t_f$, $w_u$, and $w_c$ is shown in
Fig.~\ref{disegnotape}. Introducing the effective superconducting
fraction $\eta_{eff} = (w_f t_f)/(w_u t_u)$, the last equation can
be written as:
\begin{equation}\label{eq7b}
\frac{\rho_{eff}}{\rho_m}=\frac{1-\frac{w_f}{w_u}}{1-\frac{w_f}{w_u}+\eta_{eff}}
\end{equation}

\subsection{\label{sec2:3}Theoretical values of $\chi_0$}
 For a superconductor in Meissner state, the $\chi_0$ expressions for infinite strip and
finite x-z array have been reported by Fabbricatore et al.
\cite{PRB61(2000)6413,SST13(2000)1327} .\\ \indent If our samples
are considered like a single strip with rectangular cross section,
$\chi_0$ is given by:
\begin{equation}\label{eq9}
  \chi_0=\frac{\pi w_z}{4 d_z}
\end{equation}
where $w_z$ and $d_z$ are respectively  the width and the
thickness of the filamentary zone. For a more realistic geometry
like a finite x-z  array with $2 \times 8$ filaments, the
expression for $\chi_0$ is very complicate and is given by:
\begin{equation}\label{eq10}
  \chi_0 =\chi_{0}(1)\frac{f_x}{f_z}
\end{equation}
where $\chi_0(1) = {\pi w_f}/{4 t_f}$, is the geometrical factor
calculated according to Eq. (\ref{eq9}), and
\begin{equation}\label{eq10a}
  f_x =\left[1+\frac{2}{\pi}(\beta-1)\textrm{arcsec}\left(\frac{n_x+5\beta
  -1}{5\beta}\right)\right]\textrm{;}
\end{equation}
\begin{equation}\label{eq10b}
  f_z
  =\chi_0(1)\left(\chi_0(z-\infty)+\frac{\chi_0(1)-\chi_0(z-\infty)}{n_z^{0.8}}\right)^{-1}\textrm{;}
\end{equation}
where  $n_x$ is the number of filaments in the $x$ direction and
$n_z$ is the number of filaments in $z$ direction. Moreover
\begin{equation}\label{eq11}
  \beta = \frac{\chi_0(x-\infty)}{\chi_0(1)}\frac{D_c}{w_f}\frac{d_f+D_c-w_f}{D_c-w_f}
\end{equation}
where\cite{PRB54(1996)13215}
\begin{equation}\label{eq12}
  \chi_0(x-\infty) =-\frac{2D_c^2}{\pi w_f t_f}\ln\left[\cos\left(\frac{\pi w_f}{2
  D_c}\right)\right]\textrm{;}
\end{equation}
\begin{equation}\label{eq12a}
  \chi_0(z-\infty) =\frac{2t_u^2}{\pi w_f t_f}\ln\left[\cosh\left(\frac{\pi w_f}{2
  t_u}\right)\right]\textrm{;}
\end{equation}
and $D_c$ is the distance between the centers of each column,
whereas $t_u$ is the distance between the centers of two neighbor
filaments.

\section{\label{sec3}Experimental details}

\subsection{\label{sec3:1}The samples}
\begin{figure}[htb]
\includegraphics[width=\figwidth, clip]{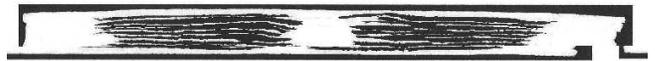}
\caption{Cross section of BSCCO/Ag sample \textbf{A}} \label{fig1}
\end{figure}

The behaviour of two different kinds of bi-columnar BSCCO(2223)
tapes has been investigated by means of the a.c. susceptibility
technique. The first BSCCO/Ag tape (named sample \textbf{A}) was
prepared with 16 filaments in pure silver matrix with a stack of 8
filaments for each column separated by about 0.3~mm of pure
silver. The external sheath is also made with the same material.
\\ The geometry of the second sample (sample \textbf{B}) is very similar
to that of the sample \textbf{A}, but the number of filaments is
15 and therefore there are 8 filaments in one column and 7 in the
other. The metallic sheet between filaments is a Ag/Mg(0.4\%)
alloy and the matrix which embeds the whole filamentary zone is a
Ag/Mg(0.4\%)/Ni(0.22\%) alloy. For both samples, strong bridging
could be present between filaments in the same column. The initial
length of the sample \textbf{A} was 61.7~mm, whereas the sample
\textbf{B} was 61.6~mm long. Preliminar transport measurements
have shown that sample \textbf{A} has a higher critical current
than sample \textbf{B}. This indicates that the grains alignment
or the grains connectivity in sample \textbf{B} is worse than in
sample \textbf{A}. The main features of the samples are summarized
in Tab.~\ref{tab1} and in Fig.~\ref{fig1} the cross section of the
sample \textbf{A} is shown. In  Tab.~\ref{tab2} the different
sizes are reported.
\begin{table}[bht]
\centering
\begin{tabular}{l c c}
\hline Sample & \textbf{A} & \textbf{B} \\ \hline matrix & Ag &
Ag/Mg(0.4\%)
\\ External sheath & Ag & Ag/Mg(0.4\%)Ni(0.22\%) \\
 fill factor ($\eta$)  &32 \% & 32\% \\
width  ($w$) & 3.3 mm & 3.1 mm \\ thickness ($t$) & 0.25 mm & 0.24
mm\\ initial length ($\ell$) & 61.7 mm & 61.6 mm
\\ critical current ($I_c$) & 23.5 A & 17.5 A\\
c. c. density ($J_c$) & 9653 A/cm$^2$ & 7173 A/cm$^2$
\\
 exponent ($n$) & 15 & 15 \\
 \hline
\end{tabular}
\caption{Main features of the samples \textbf{A} and
\textbf{B}}\label{tab1}
\end{table}

\begin{table}[tb]
\centering \vskip 1cm
\begin{tabular}{c c c c c}
\hline
 $w_c$ (mm) \ \  & $t_c$ (mm) \ \  & $D_c$ (mm) \ \  & $w_z$ (mm) \ \ & $t_z$ (mm) \ \ \\
 \hline 1.0  & 0.21  & 1.3  &  2.3  & 0.21\\ \hline
  $w_u$ (mm) & $t_u$ (mm) & $w_f$ (mm) & $t_f$ (mm) &
$t_u-t_f$ (mm) \\ \hline 1.3 & 0.027  & 1.0  & $\simeq $ 0.025 &
$\simeq$ 0.002\\
 \hline
\end{tabular}
\caption{Geometrical dimensions of the tapes as defined in
Fig.~\ref{disegnotape}}\label{tab2}
\end{table}
\indent In the samples \textbf{A} and \textbf{B}, the resistivity
of the different matrices was experimentally measured by a D.C.
transport measurement on samples without superconducting material.
For pure Ag matrix $\rho_m = 0.27 \ \mu\Omega$~cm, while the
resistivity of the Ag/Mg alloy is $\rho_m = 1.08 \ \mu\Omega$~cm.
By using the expression (\ref{eq4}) and the dimensions reported in
Tab.~\ref{tab2}, the expected ratio $\rho_{eff}/\rho_m$ is 0.25.
Therefore  the expected effective resistivity for the Sample
\textbf{A} is $0.068 \ \mu\Omega$~cm, while it is $0.27 \
\mu\Omega$~cm for the sample \textbf{B}. Finally, the estimated
value for $\eta_{eff}$ is
 71\%.

\subsection{\label{sec3:2}A.C. loss measurements}
The a.c. losses were measured by using the standard a.c.
susceptibility technique.
\begin{figure}[htb]
\includegraphics[width=\figwidth, clip]{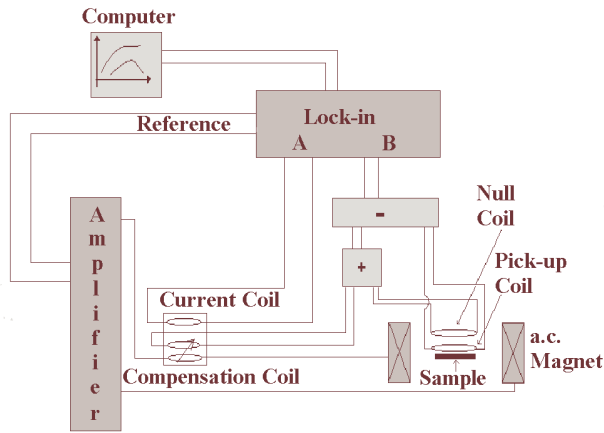}
\caption{Sketch of the experimental set-up employed for a.c.
measurements} \label{susc}
\end{figure}
An a.c. susceptometer with a system of coils suitable for
measurements on sample with length up to 12~cm is sketched in
Fig.~\ref{susc}. The susceptometer comprises a.c. electromagnet
which produces an a.c. magnetic field with $B_0$ up to 50~mT, with
a field homogeneity within $1\%$ on a 8~cm length and $2\%$ on
12~cm. The power to the a.c. magnet is delivered by an amplifier
controlled by the signal from the internal oscillator of the
lock-in amplifier. The a.c. field induces a voltage in two coils:
the pick-up coil, which is very close to the sample surface, while
the second, the null coil, which is 1~cm apart. The a.c. magnet
and both the coils are placed in a reinforced plastic cryostat, so
no eddy currents are induced in the cryostat walls. The system is
cooled by liquid nitrogen and all the measurements have been
performed at 77~K. Since the pick-up coil and the null coil are
not perfectly identical, a variable compensation system is also
used. The input signal for the lock-in is the result of the
difference between the pick-up voltage and the sum of the null and
the compensation voltages. The current which flows in the a.c.
magnet is inductively measured by the current coil. \\ \indent The
two original full length samples were subsequently cut several
times to vary the sample lengths. In each of these operations, the
tape boundaries were polished and protected with grease.
Measurements were performed in the frequency range from 1~Hz to
1000~Hz in the field amplitude ranging from 0.05~mT to 45~mT.

\subsection {\label{sec3:3}Experimental measurement of $\chi_0$}
To experimentally evaluate  the $\chi_0$ value of the samples, the
technique described in detail by Fabbricatore et al.
\cite{SST13(2000)1327} was used. By performing the measurement at
very low magnetic field amplitude (such as to assure  the Meissner
state), the constant $\chi_0$ can be determined by using the
relation:
\begin{equation}\label{eq1b}
  \chi_0 = G_c\frac{V_{coil}}{V_{sample}}\frac{U_{meas}}{U_{coil}}
\end{equation}
where $U_{meas}$ is the in input signal of the lock-in amplifier
 when the sample is placed in the pick-up coil, $U_{coil}$ is the voltage of null coil only,
$V_{sample}$ is the volume of superconducting fraction, $V_{coil}$
is the volume of the pick-up coil  and $G_c$ is a constant
depending on the geometry of the coils.

\section{\label{sec4}Experimental results}

\subsection{\label{sec4:1}Determination of $\chi_0$}
\begin{figure}[!h]
\includegraphics[width=\figwidth, clip]{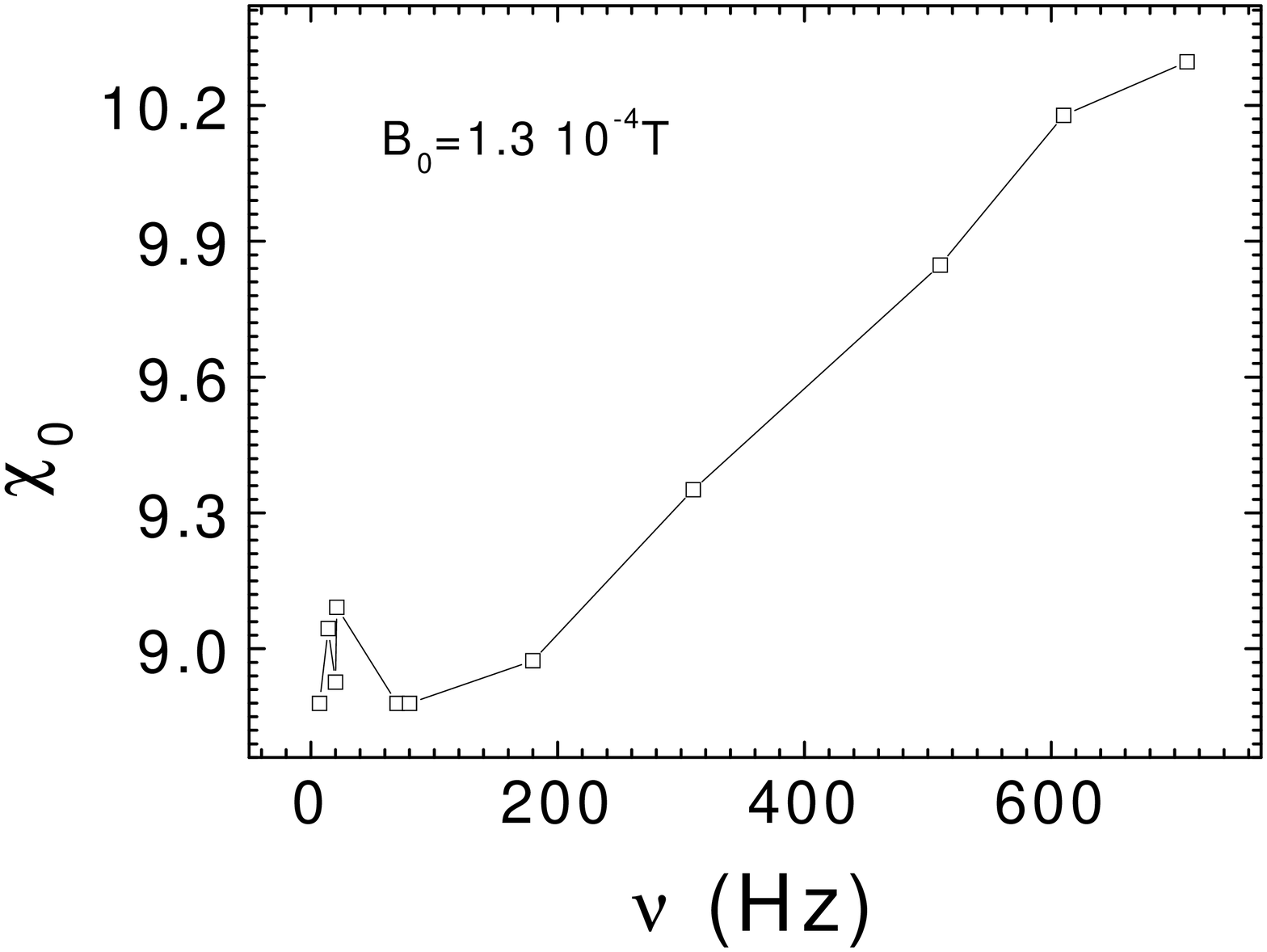}
\caption{Frequency dependence of $\chi_0$ as measured on the
sample \textbf{A} at 1.3 G} \label{chi0freq}
\end{figure}

The value of $\chi_0$ was measured on a samples of 6.4~mm length
at the frequencies of 7~Hz and 21~Hz. \\ \indent In order to check
the frequency dependence of $\chi_0$, measurements in the
frequency range between 7~Hz and 710~Hz,  reported in
Fig.~\ref{chi0freq}, were performed on the sample \textbf{A}. All
these measurements were done  at the magnetic field amplitude of
1.3~G, enough to achieve a complete shielding of the sample. At
low frequency this shielding corresponds to the Meissner state of
the separated filaments in the tape. As the frequency increases,
the shielding is achieved first in the whole filamentary zone and
then, at higher frequency, in the whole tape.
\\ \indent  From the measurements at 7~Hz, the value of $\chi_0$ for the sample \textbf{A}
 is 8.9, while it is 8.8 for the sample
\textbf{B}. This is not surprising if we consider the very similar
geometry of the two tapes. \\
\begin{table}[htb]
\centering
\begin{tabular}{l c }
\hline $\mathbf{\chi_0}$ & \\
 \hline strip $w_z \times t_z$ Eq.~(\ref{eq9}) & 9.8 \\
strip $w \times t $ Eq.~(\ref{eq9}) & 10.37
\\ x-z array Eq.~(\ref{eq10}) & 9.0
\\ Measured value sample \textbf{A} & 8.9
\\ Measured value sample \textbf{B} & 8.8\\
 \hline
\end{tabular}
\caption{Theoretical and measured $\chi_0$. }\label{tab3}
\end{table}
\indent The comparison between the theoretical and measured values
of $\chi_0$ reported in  Tab.~\ref{tab3} shows that the
experimental values are close to those theoretically calculated
for an x-z array.  Looking at the Fig.~\ref{chi0freq}, the
$\chi_0$ value is 9.0 at low frequency, it increases up to 9.9 at
510~Hz, and becomes 10.37 at 710~Hz. By comparing these values
with the theoretical values reported in Tab.~\ref{tab3}, we can
observe that the value measured at 510~Hz is closer to the one
calculated for a infinite strip with rectangular cross section, by
taking the width and thickness of the superconducting zone.
Furthermore, the value measured at 710~Hz is closer to that
calculated for a strip with the whole tape dimensions. Therefore
the behaviour of the experimental $\chi_0$, suggest that the whole
tape dimensions can be considered only at higher frequencies.
\\ \indent The interpretation is straightforward: as the frequency
increases, a larger volume of the tape
 is shielded by the
current flowing in the metallic matrix, both in the
superconducting core and in the central  metallic core. Therefore,
as the frequency increases, the effective geometry of the sample
changes from an x-z array geometry to a monofilamentary strip and
therefore, the $\chi_0$ values measured changes accordingly.

\subsection{\label{sec4:2}A.C. susceptibility measurements}
\begin{figure}[!htbp]
\includegraphics[width=\figwidth, clip]{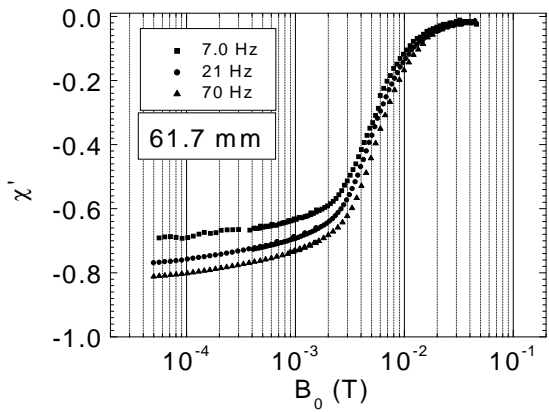}
\caption{$\chi'$  as function of the magnetic field amplitude
$B_0$, measured at different frequencies on the 61.7~mm long
sample \textbf{A} } \label{fig3}
\end{figure}

\begin{figure}[htb]
\includegraphics[width=\figwidth, clip]{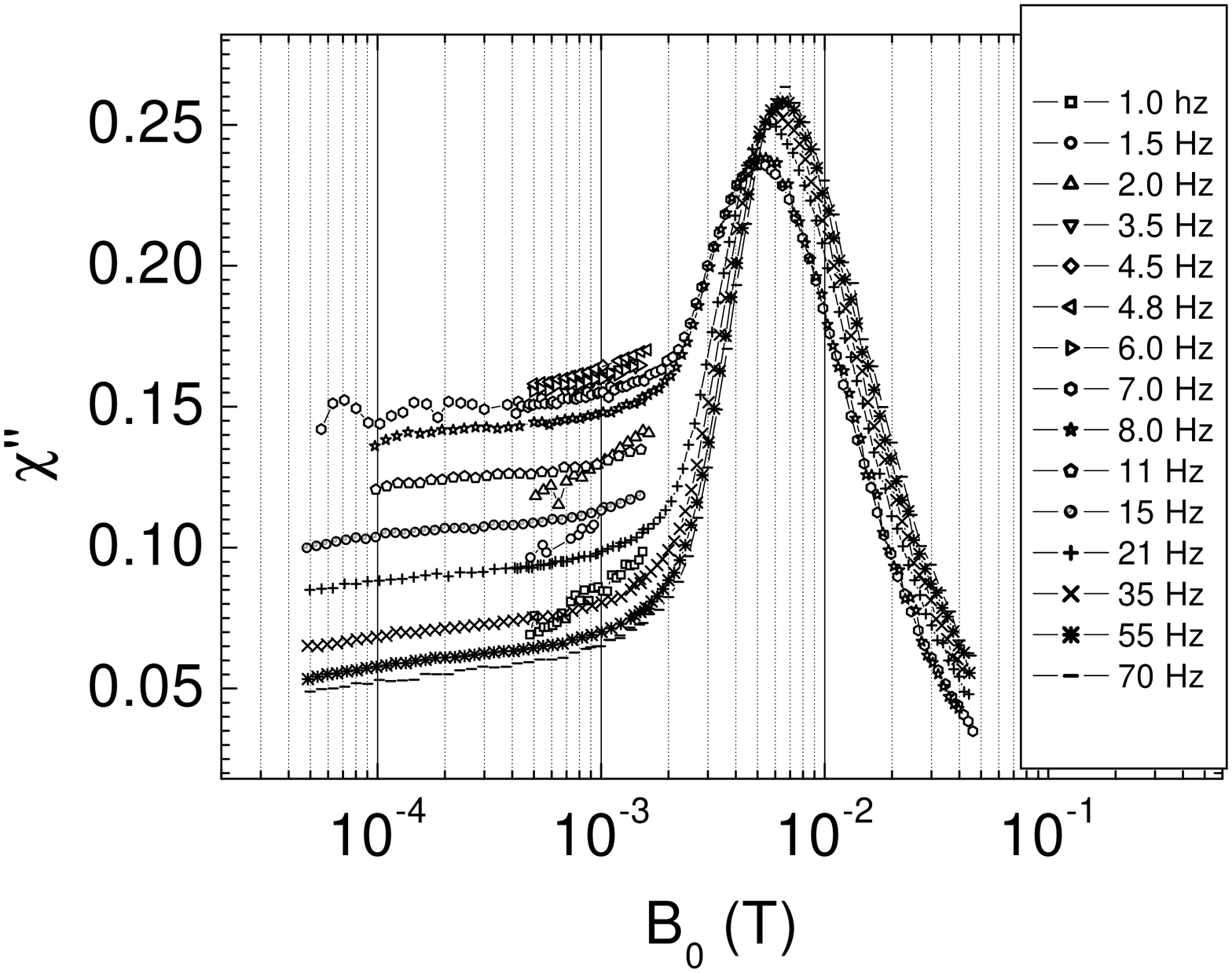}
\caption{Measurements of $\chi''(B)$ in the frequency range from
1~Hz to 70~Hz on the 61.7~mm long sample \textbf{A}. The curves
measured in the 1~Hz - 6~Hz range halt in 1.5~mT } \label{fig4}
\end{figure}

\begin{figure}[htb]
\includegraphics[width=\figwidth, clip]{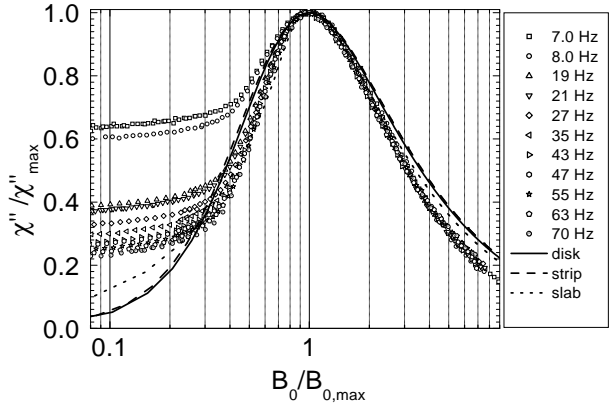}
\caption{Measurements of $\chi''(B)$ in the frequency range from
7~Hz to 70~Hz on the sample \textbf{A} with length 61.7~mm,
plotted in normalized units and compared with the analytical
results for different geometries.} \label{fig5}
\end{figure}

In Fig.~\ref{fig3} and in Fig.~\ref{fig4}, the real ($\chi'$) and
the imaginary ($\chi''$) part of the a.c. susceptibility as
function of $B_0$ are shown as measured on the 61.7~mm long sample
\textbf{A} at different frequencies.
In Fig.~\ref{fig3}, we can observe that a larger part of the tape
is screened as the frequency of the magnetic field increases. Such
frequency dependence has not been observed when $\chi_0$ was
determined on much shorter samples. Now, the coupling currents
contribute to the overall shielding and thus to $\chi'$.
Nevertheless, the correctness of experimental technique is tested
by checking that the value of $\chi'$  approaches -1 at low field
and rises up to 0 at high field.
\\ \indent In Fig.~\ref{fig4} a  set of $\chi''(B_0)$ curves,
measured in the frequency range from 1~Hz up to 70~Hz, is shown.
For frequencies ranging from 1~Hz to 6~Hz our power supply does
not allow  to reach magnetic fields higher than 1.5~mT.
 We can observe that $\chi''$
strongly depends  on the frequency. In the low field region ($B_0
\leq 1 $~mT) $\chi''$ depends slightly on $B_0$ and if we analyze
the its frequency behaviour, at fixed $B_0$, we find that $\chi''$
reaches its maximum value around 4.8~Hz.  However this maximum
value is equal to about 0.15 whereas the Campbell model predicts a
value of 0.5 (see Eq. (\ref{eq3a})). For these reason ,the
$\gamma$ factor introduced in Eq. (\ref{eq3a}) is equal to 3.3. In
the low field region, the light dependence on $B_0$ and the large
frequency dependence are in agreement with the expected behaviour
in presence of coupling losses dominant in comparison with the
filaments hysteretic losses.
\\ \indent In the high field region, the full coupling among
filaments is expected, leading to a behaviour similar to a
critical state with the typical peak of $\chi''$ in the amplitude
dependence. In this region, as the frequency rises, the maximum
value ($\chi''_{max}$) increases and shifts at higher field
amplitudes. Such increasing is due to a larger contribution of
resistive effects with respect to the critical
state\cite{PRB59(1999)11539}.
 In order to compare the
 experimental curves with the analytical results for the a.c. susceptibility,
 the data reported in Fig.~\ref{fig4}
 have been plotted in Fig.~\ref{fig5} in terms of normalized units $\chi''/\chi''_{max}$ and
$B_0/B_{0,max}$ (where $B_{0,max}$ is the magnetic field amplitude
where $\chi''$ has its maximum value).
 In the same figure, the dotted,
dashed and continuous lines are the a.c. susceptibility as
calculated in the framework of the critical state model,
respectively for a slab \cite{PRL8(1962)250,Goldfarb} , a thin
strip \cite{PRB49(1994)9024} and a thin disk
\cite{PRB50(1994)9355}. In the high field range, in particular
above $B_{0,max}$, the experimental measurements fall on a single
curve. The used analytical models do not fit very well the
experimental data. However the scaling of the data shows that the
sample behaviour, at least in the high field range, can be treated
in good approximation within the critical state model framework.
Numerical calculations of a.c. susceptibility of a superconducting
thick strip \cite{PRB54(1996)4246,PRB58(1998)6523}could give a
better agreement, also  taking into account the field dependence
of the critical current density ($J(B)$)\cite{PC372–376(2002)977}
and thermal activated creep phenomena \cite{PRB59(1999)11539}.

\section{\label{sec5}Discussion}
In this section the total losses are determined  by using the
equation~(\ref{eq1}) and discussed in the subsection \ref{sec5:1}.
In the subsection \ref{sec5:2} the losses are analyzed as function
of the frequency, at fixed magnetic field amplitudes much lower
than the full penetration field  of the composite specimen, in
order to be in the limit for the validity of the Campbell model.
Finally in the last section the effective resistivities are
determined and discussed.
\subsection{\label{sec5:1}Total losses}
\begin{figure}[htb]
\includegraphics[width=\figwidth, clip]{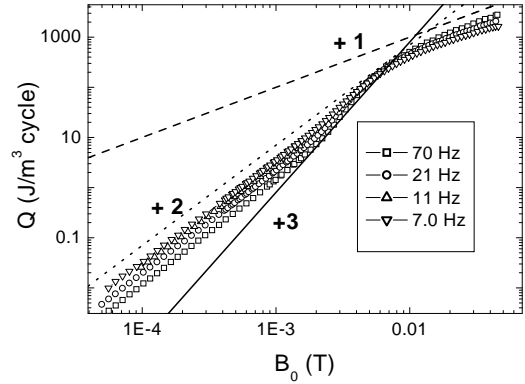}
\caption{Total losses density as function of the applied magnetic
field, measured on the sample \textbf{A} of 61.7~mm long. The
dashed and dotted lines are used to compare the experimental data
with the expected theoretical slopes  for the coupling losses
respectively in the low and the high field regions. The continuous
line is the theoretical slope as expected for hysteretic losses at
fields lower than the full penetration field.}\label{fig7}
\end{figure}
According to equation (\ref{eq3a}), in Fig.~\ref{fig7} the total
losses densities ($Q$) measured in the sample \textbf{A}, 61.7~mm
long are reported as function of $B_0$, for different frequencies.
Very similar dependencies were found for the sample \textbf{B}. In
Fig.~\ref{fig7} we also report the theoretical slopes expected for
the coupling losses in the low field (dotted line) and in the high
field (dashed line) regimes. In fact, according to Eq. (\ref{eq3})
for $B_0$ much lower than $B_{0,max}$ the coupling losses should
have a quadratic dependence on $B_0$. At higher fields, in a
critical state description,  a cubic dependence can be expected
for $B_0 < B_{0,max}$, whereas in the high field region $B_0
> B_{0,max}$ the dependence is linear. Nevertheless for $B_0 \ll
B_{0,max}$, a contribution due to the hysteretic losses in the
single filaments ( $Q_h \propto B_0^3 $) should be added to the
coupling losses. Therefore in this field region $Q \propto k_1
\times B_0^3 + k_2 \times B_0^2$ where $k_1$ and $k_2$ are two
proportionality constants and in a log-log plot the slope of the
loss density can range from 2 to 3. Our experimental data for
fields lower than 1~mT, show a slope very close to the expected
value for a pure coupling regime. Deviations from the square slope
are observed as the length of the samples is reduced. This is in
agreement with the reduction of the coupling loss density as the
sample length is reduced,  whereas the hysteretic loss density
does not change.

\subsection{\label{sec5:2}Frequency dependence of the losses at low magnetic fields}
\begin{figure}[htb]
\includegraphics[width=\figwidth, clip]{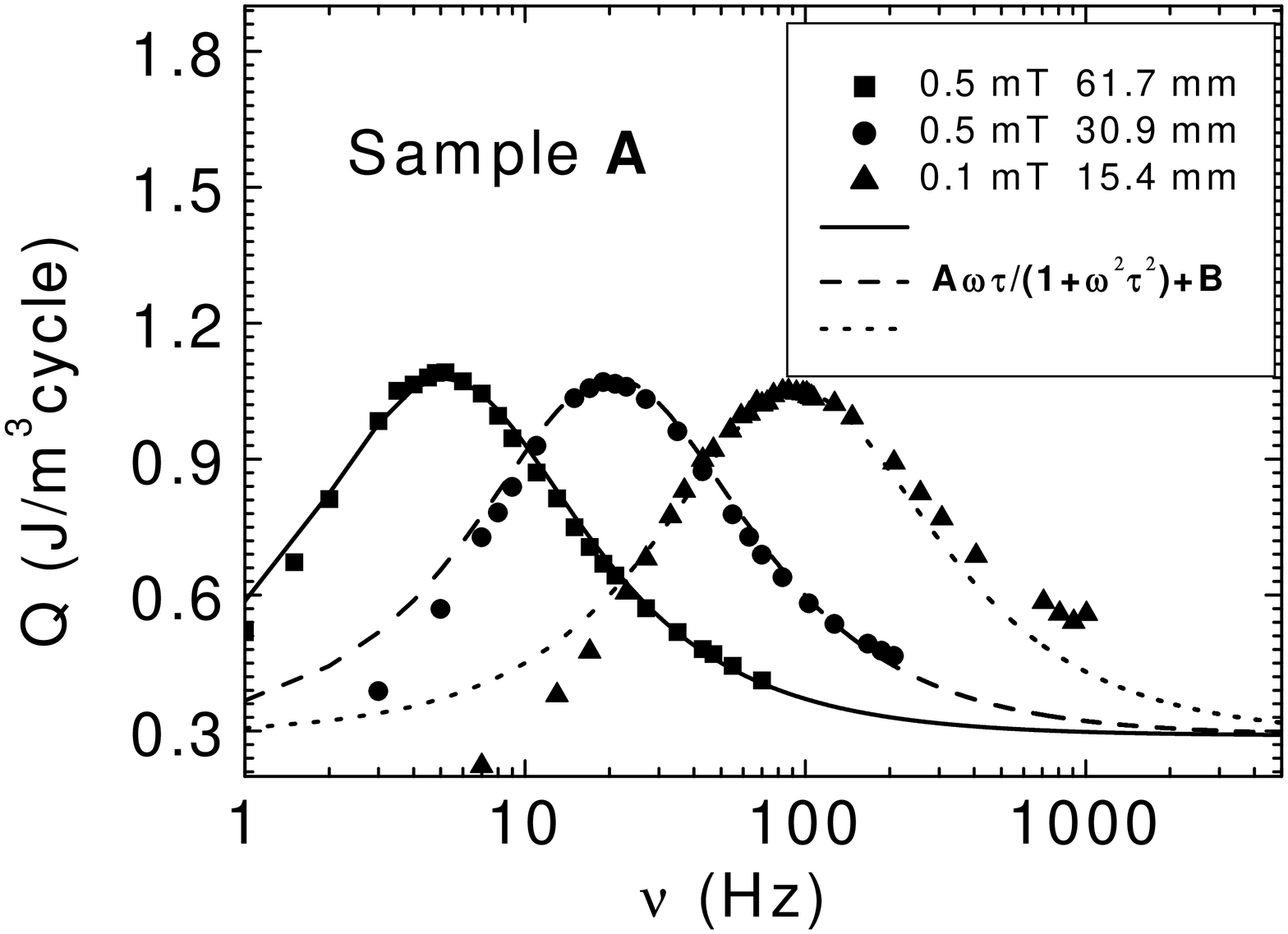}
\includegraphics[width=\figwidth, clip]{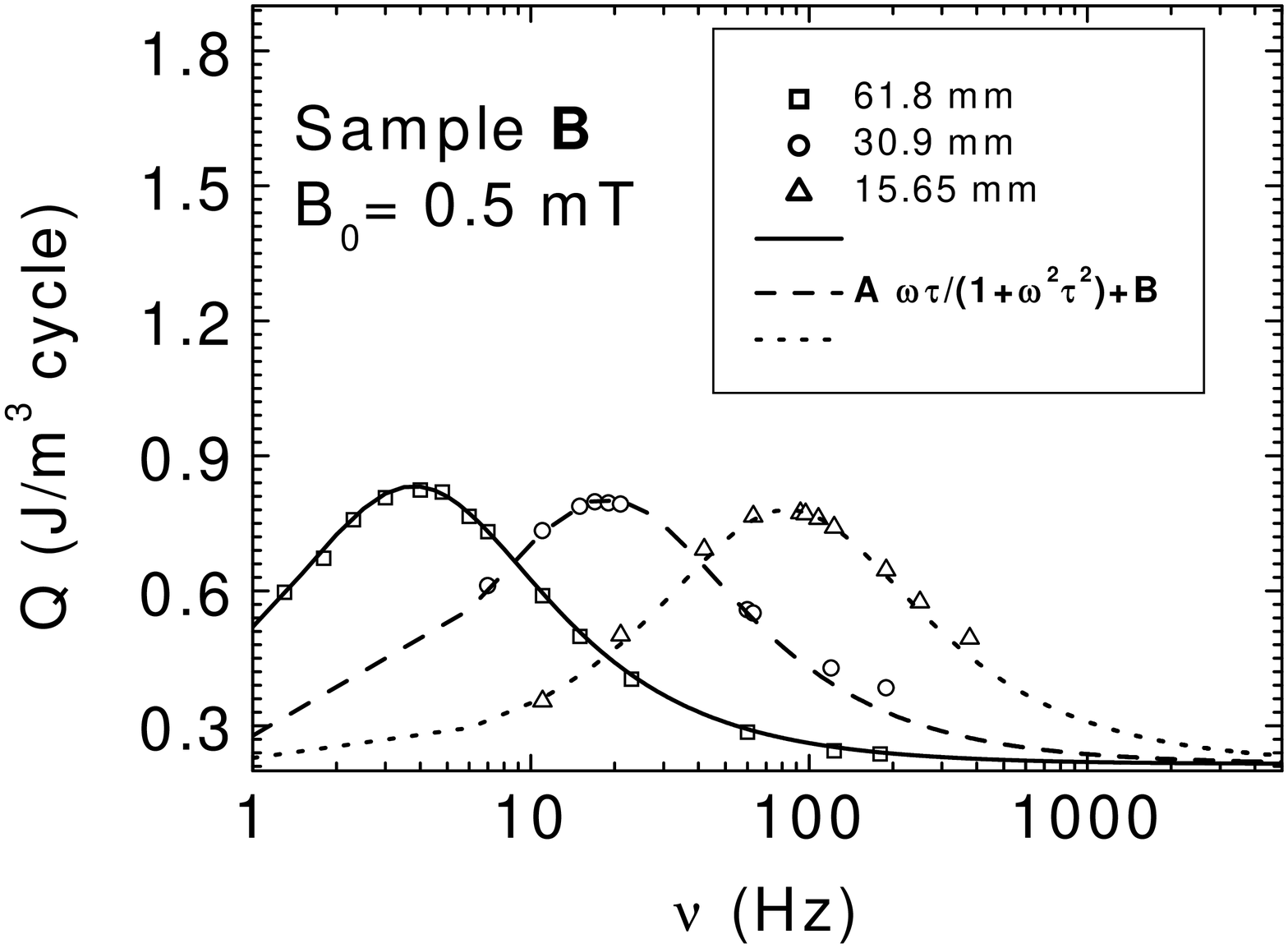}
\caption{Frequency dependence of the losses for the sample
\textbf{A} and the sample \textbf{B}, cut in pieces with different
lengths. The lines are the fit obtained with the help of the
Eq.(20).}\label{fig8}
\end{figure}

\begin{figure}[htb]
\includegraphics[width=\figwidth, clip]{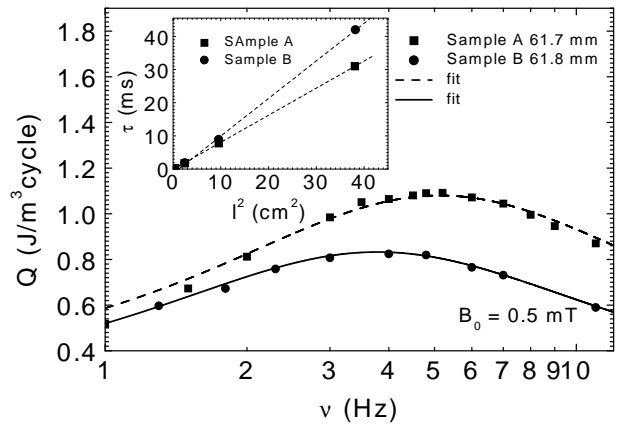}
\caption{Comparison between the frequency dependencies of the
losses for the samples \textbf{A} and \textbf{B}, in the frequency
region around their maximum. Inset: Dependence of the time
constant $\tau$ on the square of the sample length (see also
Tab.~\ref{tab4}.)}\label{fig9}
\end{figure}

Losses have been investigated as function of the frequency at
field much lower than $B_{0,max}$, by measuring several pieces
with different lengths, cut from our original samples \textbf{A}
and \textbf{B}. As reported in Fig.~\ref{fig8}, the losses exhibit
a maximum that shifts towards higher frequencies as the samples
length decreases. The experimental data are compared with the
results derived from the Campbell model \cite{Cryo22(1982)3}. In
particular, the data have been fitted by adding to the coupling
loss a frequency independent contribution ($\beta$) due to  the
filaments hysteretic loss:
\begin{equation}\label{eq14}
  Q_{fit}(\omega)=\alpha\frac{\omega \tau}{1+\omega^2\tau^2}+\beta
  \end{equation}
For each fit we have used three different parameters, which were
not free. In fact, for $\beta$  we have employed the same value in
the three different fits, performed for the samples with different
lengths. These values are respectively equal to $\beta$=0.29
J/m$^3$ cycle for the samples \textbf{A} and $\beta$=0.215 J/m$^3$
cycle for the samples \textbf{B}. Since the hysteretic losses are
proportional to the \hyphenation{cri-ti-cal cur-rent den-si-ty}
$J_c$, we compared the ratio ($J_{c,\textrm{\textbf{\footnotesize
B}}}/J_{c,\textrm{\textbf{\footnotesize A}}}$), with the ratio of
the $\beta$ values found by the fit. $J_{c,\textrm{\textbf{\tiny
B}}}/J_{c,\textrm{\textbf{\tiny A}}} = 0.743 $, that is in good
agreement with the ratio $\beta_{\textrm{\textbf{\tiny
B}}}/\beta_{\textrm{\textbf{\tiny A}}}$ = 0.741. \\ The parameter
$\alpha$ is determined by the maximum value of $Q$ which is equal
to $\alpha/2+\beta$. The $\alpha$ values are in the range ($1.52
\div 1.58$)~J/m$^3$ cycle for the sample \textbf{A} and in the
range ($1.13  \div 1.23$)~J/m$^3$ for the sample \textbf{B}. \\
The $\tau$ values, obtained from the fits, are reported in
Tab.~\ref{tab4} whereas in the inset of Fig.~\ref{fig9} its linear
dependence on the square of the sample length is shown. The linear
fit is very satisfactory for both samples.
As shown in Fig.~\ref{fig8}, the  fits for the samples \textbf{B}
are better than those for the samples \textbf{A}. The largest
deviations are observable for the sample \textbf{A} of 15.4~mm
length. In fact, for this sample, in Fig.~\ref{fig8} we show the
measured losses for a field amplitude of 0.1~mT, because at 0.5~mT
the observed discrepancy between the fit and the experimental data
is even larger. This can be ascribed to some other frequency
effect which becomes important in the frequency range
(100~-~1000~Hz) for pure Ag matrix sample.
Indeed, As pointed out by Tak\'acs
\cite{PC354(2001)202,ADVCRYOENG}, the introduction of two
different time constants, $\tau_0$ and $\tau_1$, in the expression
for the coupling losses (see formula (\ref{eq3})) is necessary in
the case of flat composite cables, leading to:
\begin{equation}\label{eq16}
  Q_c =
  \frac{\gamma B_0^2}{2\mu_0}\left[2\pi\chi_0\left(\frac{\omega\tau_0}{1+\omega^2\tau_1^2}\right)\right
  ]
\end{equation}
In Eq. (\ref{eq16}) $\tau_0$ is related to the resistivity of the
 coupling currents loops and it is determined by the zero
frequency limit of $Q(\nu)$, while $\tau_1$ is related to the mean
inductance of these loops, and it is determined by the position of
the maximum in $Q(\nu)$.
 Unfortunately, for most of the samples, at low
frequencies there are not enough data to be linearly fitted in
order to determine the value of $\tau_0$. Nevertheless, for the
15.5~mm long sample \textbf{A} this fitting procedure has been
possible, and the value for $\tau_0$ has been estimated to be 1.5
ms, which is not so different from 1.7 ms found for $\tau_1$ by
using the same data. Increasing the length of the sample, the
frequency where the maximum in $Q(\nu)$ occurs decreases and
therefore a smaller difference between $\tau_0$ and $\tau_1$ is
expected.

\subsection{\label{sec5:4}Effective Resistivity}
\begin{figure}[htb]
\centering
\includegraphics[width=\figwidth, clip]{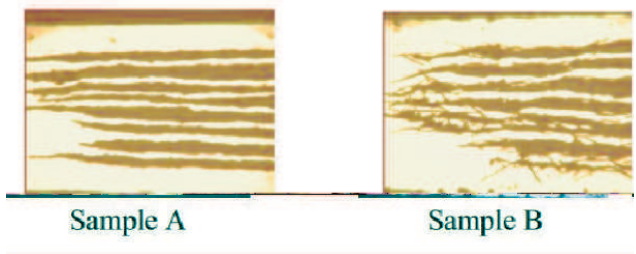}
\caption{Micro-graphs of a cross section of the sample \textbf{A}
and \textbf{B}. In the sample \textbf{B} the intergrowths are
clearly visible.} \label{fig11}
\end{figure}

By using the experimental values of $\tau$ and $\chi_0$, the
effective resistivity of the metallic matrix has been determined.
\\ For samples with the same length and the same geometry the
relation $\tau_{\textrm{\textbf{\tiny
A}}}/\tau_{\textrm{\textbf{\tiny B}}} =
\rho_{\textrm{\textbf{\tiny B}}}/\rho_{\textrm{\textbf{\tiny A}}}$
is valid. If $\rho_{\textrm{\textbf{\tiny A}}} <
\rho_{\textrm{\textbf{\tiny B}}}$ we should have
$\tau_{\textrm{\textbf{\tiny A}}} > \tau_{\textrm{\textbf{\tiny
B}}}$, which means that the frequency ($\nu_{max}$) where the
maximum in the $Q_c(\nu)$ occurs  is lower for the sample with
lower resistivity. The directly measured matrix resistivity of the
sample \textbf{A} is 4 times smaller than that for the sample B
(see section \ref{sec3:1}). Since the structure of the samples is
very similar (as confirmed by the experimentally found values of
$\chi_0$), we expect ($\nu_{max,\textrm{\textbf{\tiny
A}}}<\nu_{max,\textrm{\textbf{\tiny B}}}$). It is striking to see
that the experimental data in Fig.~\ref{fig9} demonstrate the
opposite behaviour.  The obtained results are summarized in
Tab.~\ref{tab4}.
\begin{table}[tb]
\centering
\begin{tabular}{c c c c }
\hline $\ell$ (mm)  &  $\tau$ (ms)  & $\chi_0$ & $\rho_{eff}$
($\mu\Omega$~cm)\\ \hline sample \textbf{A}
\\ \hline 61.7 &  31 & 8.9 & 0.176 \\ 30.9 &
 7.8 & 8.9 & 0.175 \\ 15.5  & 1.75 &
& 0.192
\\ 6.5  & 0.32 & 9.8 & 0.172 \\ \hline sample \textbf{B}
& & & \\ \hline 61.8  & 42 & 8.8 & 0.132 \\ 30.9 & 8.9 & 8.8 &
0.155 \\ 15.6  & 1.9 & 9.0 & 0.182\\ \hline
\end{tabular}
\caption{Values of the quantities $\tau$, $\chi_0$ and
$\rho_{eff}$ as determined from the experimental data for all the
considered samples.}\label{tab4}
\end{table}
For the sample \textbf{A}, the value of the effective resistivity
is 3 times higher than the value expected from Eq.~(\ref{eq7}).
Moreover the effective resistivity found for the sample \textbf{B}
is lower than both the $\rho_{eff}$ of the sample \textbf{A},  and
 its $\rho_{eff}$ determined by Eq. (\ref{eq7}). This can be
understood if the structure of the sample is considered. In
Fig.~\ref{fig11} the micrographs of a section of both the samples
\textbf{A} and \textbf{B} are shown. In the sample \textbf{B} many
intergrowths in the metallic matrix are visible. Intergrowths
between columns can generate paths with lower resistivity for the
flowing of the coupling currents. The presence of intergrowths
reduces or cancels the advantage coming from the use of a matrix
with higher resistivity in order to decrease the coupling losses.
\\ \indent On the contrary, the sample \textbf{A} shows a value of
$\rho_{eff}$ which is higher than the expected one. This high
value can be due to a contact resistance  between the
superconducting filaments and the metallic matrix.

\section{\label{sec6}Conclusions}
In this work we have studied the a.c. coupling losses on two
different sets of bicolumnar BSCCO tapes. \\  \indent For the
geometrical factor $\chi_0$, we used the experimental value
instead of the demagnetizing factor $(t_z+w_z/t_z)$ used in
literature\cite{Cryo39(1999)829,PC355(2001)325,PC370(2002)177}
which should give  the found value 13.5 instead of the value 8.9,
for both samples. Our experimental value is in agreement with the
theoretical value computed for an x-z finite array.   In this way
the main physical quantity involved in the coupling losses have
been experimentally estimated. \\ For increasing frequencies the
measured $\chi_0$ approaches the value calculated for a
superconducting strip with the  width and the thickness equal to
the dimensions of the filamentary zone of the tape. By further
frequency increasing, $\chi_0$ approaches the value calculated for
a superconducting strip  with dimensions equal to those of the
full tapes. \\ \indent  A.C. susceptibility and losses, measured
as function of the magnetic field amplitude and frequency, confirm
that the coupling losses dominate in these samples over the
hysteretic losses of the single filaments. The frequency
dependence of the experimentally measured losses at low field has
been analyzed by using the Campbell model and a good agreement
between the model and the experimental data has been obtained. The
time constants $\tau$ of the tape has been determined for
different lengths of the sample, finding, as expected, a linear
dependence of $\tau$ on the square of the sample length. The final
task has been the experimental evaluation of the effective
resistivity. In the sample \textbf{A} with a pure silver matrix,
the high value measured for $\rho_{eff}$ is interpreted as bad
electrical contacts between the filaments and the metallic matrix.
In the sample \textbf{B}, the experimentally found value
$\rho_{eff}$ lower than the expected one in an ideal sample, is
explained with the presence of intergrowths. This particular
result suggests that in this kind of samples, the enhancement of
the effective matrix
resistivity does not reduce intrinsically the coupling losses. \\
\indent Finally it is worth to point out that for low magnetic
field amplitudes, the Campbell model can be successfully used also
for BSCCO tapes containing few flat filaments. This can be used as
a starting point to analyze all the other factors which influence
the losses in more complex geometries.

\section{Acknowledgements}
We thank J\'{a}no \v{S}ouc for helpful discussion and Juraj
Tan\v{c}\'{a}r for his technical support.
\bibliography{couplinglosses}

\end{document}